\documentclass[twocolumn,prl]{revtex4}
\newcommand{\figurewidth}{84mm}

\usepackage{graphicx}
\usepackage{dcolumn}
\usepackage{amsmath}

\newcommand{\BE}{\begin{equation}}
\newcommand{\EE}{\end{equation}}
\newcommand{\BEN}{\begin{eqnarray}}
\newcommand{\EEN}{\end{eqnarray}}

\newcommand{\kms}{\,\rm km\,s^{-1}}
\newcommand{\gcc}{\,\rm g\,cm^{-3}}
\newcommand{\eV}{\,\rm eV}
\newcommand{\Al}{{\rm Al}}
\newcommand{\p}{{\rm p}}
\newcommand{\s}{{\rm s}}
\newcommand{\K}{{\, \rm K}}

\begin{document}

\title[Short Title]{
    Calculation of a Deuterium Double Shock Hugoniot from
                    Ab~initio Simulations}

\author{B. Militzer}
\affiliation{Lawrence Livermore National Laboratory,
         University of California, Livermore, CA 94550}
\author{D. M. Ceperley}
            \affiliation{Department of Physics,
             National Center for Supercomputing Applications,
             University of Illinois at Urbana-Champaign, Urbana, IL 61801}
\author{J. D. Kress}
\author{J. D. Johnson}
\author{L. A. Collins}
\author{S. Mazevet}
\affiliation{Theoretical Division, Los Alamos National Laboratory, Los Alamos, New Mexico 87545}

\date{\today }

\begin{abstract}
We calculate the equation of state of dense deuterium with two
{\it ab~initio} simulations techniques, path integral Monte Carlo and
density functional theory molecular dynamics, in the density range
of $0.67 \leq \rho \leq 1.60 \gcc$. We derive the double shock
Hugoniot and compare with the recent laser-driven double shock
wave experiments by Mostovych et al.~\cite{Mo00}. We find excellent
agreement between the two types of microscopic simulations but a
significant discrepancy with the laser-driven shock measurements.
\end{abstract}

\maketitle


Single-shock laser experiments on liquid deuterium have probed the
equation of state (EOS) up to 340~GPa~\cite{nova} and have measured a
significantly higher compressibility than predicted by standard models
like Sesame~\cite{Ke83} and by more recent experiments using magnetic
drives~\cite{Knudson}. The laser-driven experimental findings also
disagree with results from first principles
simulations~\cite{MC00,dftlanl,dftother} and leave only some chemical
models in agreement~\cite{Ro98,SC92}.  The EOS is of fundamental
importance to our understanding of many physical phenomena such as the
Jovian planets, inertial confinement fusion, and pulsed-power produced
plasmas.

In the recent laser-driven double shock (DS) experiments by Mostovych et
al.~\cite{Mo00}, final pressures of 100-600 GPa have been reached.
The results also disagree with the Sesame EOS but agree well with the
linear-mixing model~\cite{Ro98}.
Both models are based on an approximate free
energy function constructed from known theoretical limits, e.g., Sesame uses
Saha and Thomas-Fermi-Dirac theories for the electronic properties and
hard sphere reference systems to incorporate atoms and molecules.
The double shock experiments appear to provide an independent
confirmation of both the laser-driven single shock results~\cite{nova}
as well as some of the chemical models. Furthermore the findings were
interpreted as an indication that the dissociation of molecules makes
deuterium more compressible than predicted by microscopic simulations.

In this paper, we compare the Mostovych et al.~\cite{Mo00} DS results
with {\it ab~initio} simulations from path integral Monte Carlo (PIMC)
and density functional molecular dynamics (DFT-MD).  We give details of
the calculation of the secondary shock hugoniot curve from the
simulation results. In this calculation, we use either the PIMC results
or combine PIMC and DFT data.  Both methods yield very similar results.
We discuss whether the DS results indicate an increased compressibility
for the primary hugoniot curve and therefore support the single shock
experiments, despite the different temperature and densities being
reached. Furthermore, we discuss whether the process of molecular
dissociation is important.


PIMC~\cite{pimc_ceperley,MC01} is a first principles simulation method
that works directly with electrons and protons. Except for the problem
of fermi statistics, it is an exact solution of the many-body quantum
problem for a finite system in thermodynamic equilibrium. We have
treated fermi statistics by restricting the paths in the region of
positive density matrix by using a variational density
matrix~\cite{MP00} consisting of a set of Gaussian single particle
density matrices; a procedure that leads to lower free energy than the
restrictions previously employed with free particles.  The PIMC method
quantitatively describes well the phenomena of electron correlation,
formation of atoms and molecules, and temperature effects. The effect
of fermi statistics is only important well below the fermi
temperature, thus we have confidence in the results for temperatures
greater than $20\,000\K$~\cite{MC00}. The PIMC results become
increasingly accurate for higher temperatures.

For lower temperatures, where the
effects of electronic excitations are less important,
the DFT results become more reliable (relative to PIMC).
We performed fixed--volume molecular dynamics simulations, employing
finite-temperature DFT using the Mermin~\cite{Mermin} functional
with the electron and ion-kinetic temperatures set equal
and the generalized gradient approximation~\cite{perdew} (GGA).
Our study used the VASP plane--wave pseudopotential
code, which was developed at the Technical University of
Vienna~\cite{vasp1}. This code implements the Vanderbilt ultrasoft
pseudopotential scheme~\cite{vanderbilt,kresse}, and the Perdew-Wang 91 parameterization of
GGA~\cite{perdew}.


In both shock experiments, the initial state of liquid deuterium is at
a density of $\rho_0=0.171\gcc$ at atmospheric pressure and close to
zero temperature (20K). In our calculation, we match these conditions
by setting $E_0=-15.886\eV$ per atom~\cite{KW64}, $P_0=0$ and using
$n_0 = \rho_0/m$ where $m$ is the deuteron mass.
If a shock wave with shock velocity $u_\s$ and particle velocity
$u_\p$ is driven through the sample at state $(E_0,n_0,P_0)$, the new
state on the other side of the shock front $(E_1,n_1,P_1)$ follows
from conservation of mass, momentum and energy~\cite{Ze66}:
\begin{eqnarray}
\label{hug_p}
P_1-P_0 &=& m \, n_0 \, u_\s \, u_\p~, \\
\frac{n_1}{n_0} &=& \frac{u_\s}{u_\s-u_\p}~,\\
\label{hug}
0 &=&
(E_1-E_0) + \frac{1}{2}\,\left(\frac{1}{n_1}-\frac{1}{n_0} \right)\,(P_1+P_0)~.
\end{eqnarray}
For a given EOS, $u_\s$ and $u_\p$ can be derived from, $\,u_\p^2 =
\xi/\eta$ and $\,u_\s^2 = \xi \eta$ with $\xi=(P_1-P_0)/n_0 m$ and
$\eta=1-n_0/n_1$.
\label{section_single}

Before discussing the DS experiments, we will briefly review
the comparison with results for the principle Hugoniot from laser
shock wave experiments~\cite{nova} with {\it ab~initio} simulations. The
Hugoniot curves from PIMC~\cite{MC00} and DFT-MD~\cite{dftlanl,dftother} 
are in
fairly good agreement but differ substantially from the laser-driven 
experimental results~\cite{nova}, 
which exhibit a significantly increased compressibility (up
to  $6\rho_0$) compared to predictions
based on empirical models such as Sesame~\cite{Ke83}. The differences between
the experimental findings and {\it ab~initio} simulations can be
characterized by the extra internal energy per particle $E$ and
pressure $P$ needed to shift the PIMC hugoniot curve to obtain
agreement with the experimental results (see Fig. 1 in ~\cite{MC00}).
The relationship of the two shifts follows from Eq.~\ref{hug}: $\delta
E = \frac{1}{2} ( 1- n_1/n_0) \, \delta pv$ where $pv
= P/n$.
The shift needed  is $\delta E=3\eV$ per atom to the internal energy or
$\delta pv=-2\eV$ to the pressure. Both shifts are far outside the 
{\it ab~initio} error bars (order of $0.3\eV$).

In the DS experiments, a laser is used to drive a shock wave
through the deuterium sample. This primary shock propagates through
the sample and then reflects off an aluminum witness plate. This
drives a shock wave through the aluminum plate and causes a secondary
shock wave traveling in the opposite direction through the already
shocked deuterium sample. We label the final DS state of
the deuterium as $(E_2,n_2,P_2)$.

Shock waves in aluminum have been studied intensively~\cite{Mi83} and
the hugoniot is well known. We use the linear fit formula,
\begin{eqnarray}
u_\s = \left\{
\begin{array}{cl}
1.339~ u_\p +5.386~ \kms & ~{\rm if} ~u_\p < 7.557 \kms\\
1.1484 u_\p +6.8263 \kms & ~{\rm otherwise.}
\end{array}
\right.
\label{al_hug}
\end{eqnarray}
Using the initial aluminum density of $\rho_{\Al}=2.71\gcc$ and
Eq.~\ref{hug_p}, we can relate the particle velocity $u_{\p_\Al}$ to
the pressure in the shocked aluminum $p_{\Al}$.
Since the double shocked deuterium material and the shocked aluminum
are in direct contact, the pressure as well as the particle velocity
must be constant across the interface, $u_{\p 2}=u_{\p_{\Al}}$ and
$p_2 = p_{\Al}$ (impedance match). We assume that the uncertainty of
the aluminum hugoniot (Eq.~\ref{al_hug}) is approximately 1\%,
which corresponds to a negligible (2 GPa) error in the final shock 
pressure.

To study the propagation of the secondary shock front in deuterium,
one preferably goes into the frame of the shocked deuterium moving
with $u_{\p_1}$ ($\tilde{u}_{\s_2} = u_{\s_2} - u_{\p_1}$ and
$\tilde{u}_{\p_2} = u_{\p_2} - u_{\p_1}$).  In this frame, 
Eqs.~\ref{hug_p}-\ref{hug} relate the primary shock state
$(E_1,n_1,P_1)$ to the secondary shock state $(E_2,n_2,P_2)$.

To compare with the DS experiments~\cite{Mo00}, we need to
obtain the secondary hugoniot for a given EOS. Assuming a primary
shock pressure $P_1$, we calculate the first shock state
$(E_1,n_1,P_1)$ and consequently $u_{\p_1}$. Then we determine the
remaining variables as a function of $P_2$ and $u_{\p_2}$. They need
to satisfy two equations: the aluminum state has to be on the
aluminum hugoniot and the deuterium states 1 and 2 have to
satisfy the hugoniot relation Eq.~\ref{hug}.
To solve the two equations for the two unknowns, we use a Newton
procedure beginning with an initial guess. 

\begin{figure}[htb]
\includegraphics[angle=270,width=\figurewidth]{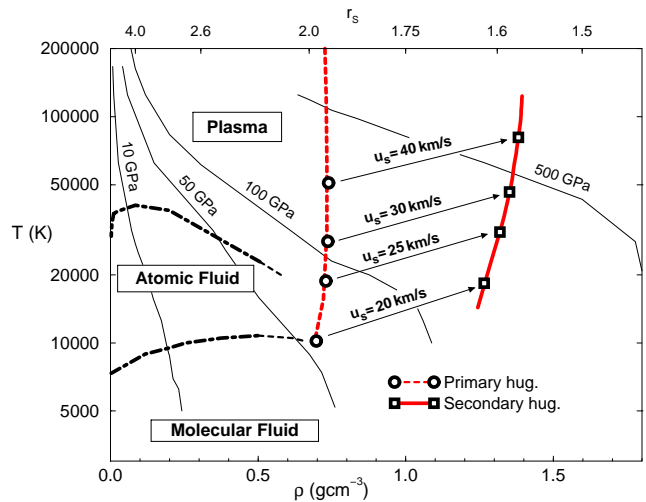}
\caption{Phase diagram of deuterium~\cite{MC01} with primary and
         secondary Hugoniot curves. The primary shock velocities are
         indicated on the arrows connecting the different shock states
         from Tab.~\ref{tab1}. The solid lines are isobars.}
\label{phase}
\end{figure}

Calculating the DS hugoniot curve in Fig.~\ref{phase}
requires an accurate EOS in the temperature and density range of $10000
\leq T \leq 100000\,\rm K$ and $0.67 \leq \rho \leq 1.4 \,\rm
g\,cm^{-3}$. 
This overlaps with the region where both PIMC and DFT simulations have
been applied.

\begin{figure}[htb]
\includegraphics[angle=0,width=\figurewidth]{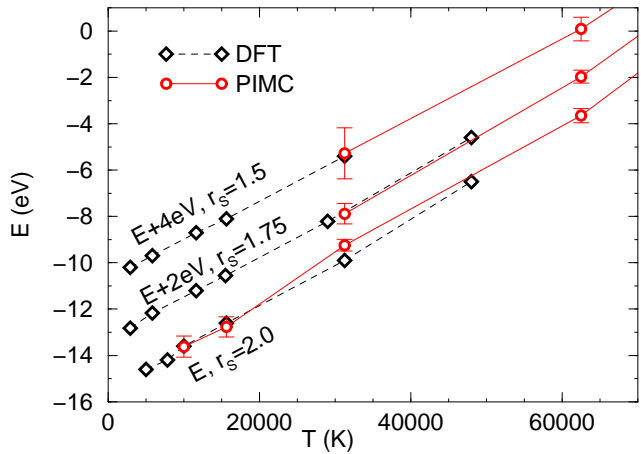}
\caption{Internal energy per atom vs temperature calculated with PIMC and DFT simulations.
         The curves for $r_s=1.75$ and 1.5 were offset by 2 and 4 eV 
	for clarity.}
\label{E_vs_T}
\end{figure}

\begin{figure}[htb]
\includegraphics[angle=0,width=\figurewidth]{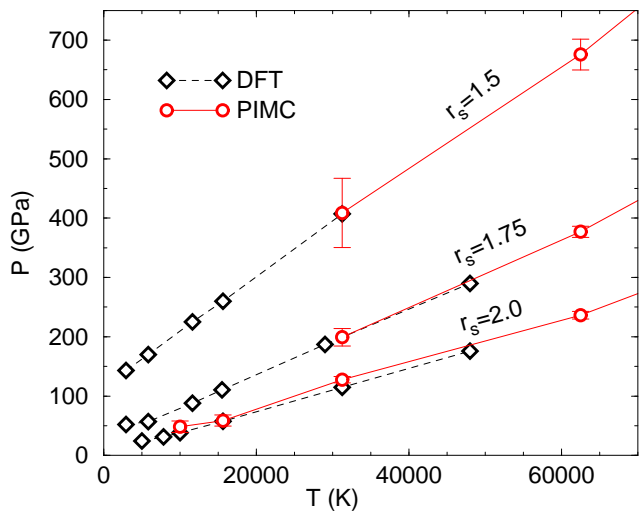}
\caption{Pressure vs temperature calculated with PIMC and DFT simulations.}
\label{P_vs_T}
\end{figure}

\begin{table}
\caption{Secondary shock pressure as a function of primary shock
         velocity comparing measurements with simulation results. 1
         and 2 label the method used for the primary and secondary
         shock states. $^*$ indicate our most reliable results and
         $^\dagger$ where the DFT EOS ~\cite{dftlanl} needed to be
         extrapolated to higher temperatures. Statistical error estimate
         in parentheses.}
\begin{tabular}{c|c|c|c|c|c}
                   & Experi-       & PIMC~1        & PIMC~1        & DFT~\cite{dftlanl}~1 & DFT~\cite{dftlanl}~1 \\
$u_{\s_1}$         &ment\cite{Mo01}& PIMC~2        & DFT~~2        & PIMC~~~2             & DFT~~~~~2 \\
$(\rm km\,s^{-1})$ &$P_2 (\rm GPa)$&$P_2 (\rm GPa)$&$P_2 (\rm GPa)$&$P_2 (\rm GPa)$       & $P_2 (\rm GPa)$\\
\tableline
20                 &               &  200(10)         &  193(6)$^*$   &    231(13)              & 223(1)\\
25                 &  400(80)      &  307(8)$^*$      &  307(6)~      &    327(7)~              & 328(1)\\
30                 &  590(90)      &  429(7)$^*$      &               &    439(4)~   \\
35                 &               &  570(7)$^*$      &               &    566(3)$^\dagger$   \\
40                 &               &  730(5)$^*$      &               &    709(3)$^\dagger$   \\
\end{tabular}
\label{tab1}
\end{table}

Our novel EOS results calculated with PIMC and DFT simulations for the
densities corresponding to $r_s=1.75$ and 1.5 are shown in
Figs.~\ref{E_vs_T} and \ref{P_vs_T} combined with earlier results
for $r_s=2.0$~\cite{MC00,dftlanl}. The overall agreement of the two
{\it ab~initio} methods is quite good.  The DFT results, more accurate at
low temperatures, smoothly join onto the PIMC results at higher
temperatures for each density.  Thus we can consider combining both
methods in different ways in the DS calculation
(Tab.~\ref{tab1}). For high shock velocities, we used the PIMC EOS
for both shock states. This yields the most reliable secondary
Hugoniot curve except for very low shock velocities such as $u_s=20
\kms$, we can also combine a primary state from PIMC with a secondary
state from DFT.  Alternatively, we can replace the PIMC data with the
DFT EOS~\cite{dftlanl} for the first shock state.

\begin{figure}[htb]
\includegraphics[angle=0,width=\figurewidth]{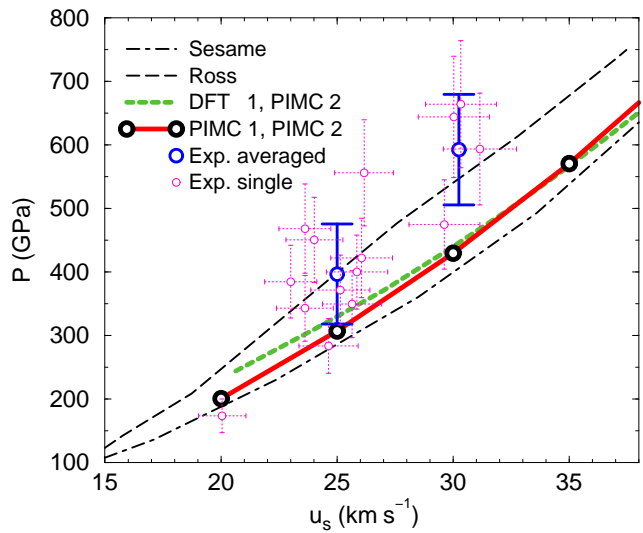}
\caption{Comparison of experimental and several theoretical
         double shock Hugoniot curves showing the secondary shock
         pressure as function of the primary shock velocity.}
\label{hug2}
\end{figure}

All different ways of combining the EOS from microscopic simulations
lead to very similar results compared to the DS
measurements~\cite{Mo00} as shown in Tab.~\ref{tab1} and
Fig.~\ref{hug2}.  {\it Ab initio} methods predict a secondary shock pressure
about 25\% lower than measured experimentally. This is outside of the
experimental error bars reported in~\cite{Mo01} (which have increased
compared to~\cite{Mo00}). Only for $u_{s_1}=25\kms$ is agreement 
found.

Both {\it ab initio} methods have statistical and finite size
uncertainties. The statistical errors, larger for PIMC, are given in
Tab.~\ref{tab1}. Finite size errors are more difficult to estimate due
to the computational demand. However, very small finite-size errors,
corresponding to only a 2\% uncertainty of $P_2$ were determined from
other DFT simulations~\cite{dftother}. In addition, PIMC results~\cite{MC01}
indicate that the finite-size dependence can not explain the difference with
single shock experiments. We estimate that both types of uncertainties
are significantly smaller than the deviations from the experiments.

The following analysis is based on the 15 individual shock
measurements, which have been reported and shown in Fig.~\ref{hug2}.
Using a $\chi^2$ computation leads to $\chi^2=39$ for the agreement
with PIMC simulations (for both first and second shocks) and
$\chi^2=16$ for the Ross model~\cite{Ro98}.  We note that if the
experimental error bars were multiplied by 1.5 then one would obtain a
reasonable agreement with the microscopic simulations.
As described for the single shock experiments, one can quantify the
differences between simulation and experiments by estimating shifts
$\delta E$ and $\delta pv$. One finds that a shift of $\delta E=(4.0
\pm 1.5)\eV$ or $\delta pv=-(3\pm 1) \eV$ needs to be applied to both
shock states in order to bring the PIMC hugoniot up in pressure to
match the measurements.

We should clarify one important point.  All of the approaches discussed
in the present work (Sesame, PIMC, and DFT) contain dissociation within
the formulation; the latter two include intermediate states and
short-lived species in the dense fluid at a sophisticated quantum
mechanical level. These states are part of the partition function and
are therefore included in PIMC. The DFT-MD method allows one to study
the dynamics of dissociation and estimate life-times~\cite{dftother}.
Therefore, the reason for the softening of
the principal Hugoniot and the behavior of the Mostovych et al.
double shock experiment cannot arise primarily from the dissociation
of molecular hydrogen as suggested in~\cite{Mo00} where the comparison
of different approximate free energy models had lead to this
conclusion.


In conclusion, 
we observed significant differences when comparing our {\it ab~initio}
simulation results with the averaged DS measurements
in~\cite{Mo01}. The discrepancies are significantly larger than the
combined error bars from the experiment and the simulation. As in the
case of the single shock experiments, {\it ab~initio} methods predict
results which are relatively close to the Sesame model~\cite{Ke83}.
When comparing with the individual shock data points, we find
agreement within the error bars only at the lower pressures.
The single shock experiments~\cite{nova} suggested that deuterium is
significantly more compressible than predicted by {\it ab~initio}
simulations. A substantial increase in internal energy or decrease in
pressure would be necessary to bring the {\it ab~initio} EOS in agreement
with these measurements. We find that comparable shifts in energy or
pressure are required to reach agreement between the DS
experiments~\cite{Mo00} and microscopic simulation results. Therefore,
it can be concluded that both types of laser-driven experiments are in
relative agreement with each other and disagree with {\it ab~initio} methods.
Clearly, more theoretical and
experimental work is necessary to resolve the discussed
discrepancy. 

\begin{acknowledgments}
Work performed under the auspices of the U.S. Department of Energy
though the University of California at the Lawrence Livermore National
Laboratory (W-7405-Eng-48, CSAR subcontract B341494) and at the Los
Alamos National Laboratory (W-7405-Eng-36).  We acknowledge useful
discussions with Dr. M. Knudson.

\end{acknowledgments}


\end{document}